\documentclass[preprint,aps,prd,showpacs]{revtex4-1}
\usepackage{amssymb}%
\usepackage{graphicx}
\usepackage{subfigure}
\usepackage{amsmath}
\usepackage{bm}
\usepackage{relsize}
\RequirePackage{xspace}
\begin{document}


\title{
 $J / \psi$ electromagnetic production associated with light hadrons at $B$ factories
}

\author{Yu-Jie Zhang $^{(a,d)}$, Bai-Qing Li $^{(b)}$, and Kui-Yong Liu $^{(c)}$}
\affiliation{ {\footnotesize (a)~Key Laboratory of Micro-nano
 Measurement-Manipulation and Physics (Ministry of Education) and School of Physics, Beihang University,
 Beijing 100191, China}\\
  {\footnotesize (b)~Department of Physics, Huzhou Teachers College, Huzhou
313000, China
}\\
  {\footnotesize (c)~Department of Physics, Liaoning University, Shenyang 110036
, China}\\
{\footnotesize (d)~Kavli Institute for Theoretical Physics China, CAS, Beijing 100190, China}}





\begin{abstract}
The electromagnetic productions of $J / \psi$ associated with light
hadrons(LH) and leptonic pairs($\mu^+\mu^-$, $\tau^+\tau^-$) at $B$
factories are studied. We find that the direct electromagnetic
production cross sections of $J / \psi$($\psi(2S)$) associated with
light hadrons is about $0.10(0.04)$~pb. The direct production cross
sections of $J / \psi$($\psi(2S)$) associated with $\mu^+\mu^-$ is
about $0.056(0.020)$~pb. If we include the contributions from
$\psi(2S)$ decay, we can get the prompt cross section $\sigma[e^+e^-
\to J/\psi +\mu^+\mu^-]=(0.068 \pm 0.002)$~pb, about $(16 \pm 5)\%$
of the Belle data $\sigma[e^+e^- \to J/\psi +X_{non-c\bar
c}]=(0.43\pm 0.09\pm 0.09)$~pb, meanwhile the $e^+e^- \to J/\psi
+\tau^+\tau^-$
 process only contributes $2 \%$. The prompt cross section $\sigma[e^+e^-
\to J/\psi +Light\ Hadrons]=(0.121 \pm 0.005)$~pb is  about $(28 \pm
8)\%$ of the Belle data.

\end{abstract}
\pacs{13.66.Bc, 12.38.Bx, 14.40.Pq}

\maketitle


\section{Introduction}
Heavy quarkonium physics is an important ground to test quantum
chromodynamics (QCD) both perturbatively and non-perturbatively.
Non-relativistic quantum chromodynamics (NRQCD) factorization
approach\cite{Bodwin:1994jh} has achieved a series of successes in
describing heavy quarkonium production and annihilation decay.
However, there are still some predictions which are less
satisfactory. For more details, see a concise
review\cite{Bodwin2002review}. Among the problematic comparisons
with experiment, the large discrepancy between theoretical
predictions and experimental data on the charmonium production in
$e^+e^-$ annihilation has drawn much attention recently.

In recent years, the B factories has provided systematic
measurements on charmonium production. Some results are puzzling,
because of the large gap between the measurements and the
theoretical predictions. For example, the cross section $\sigma(e^+
+ e^- \rightarrow J/\psi + \eta_c)$, measured by Belle
Collaboration\cite{Abe:2002rb} and Babar
Collaboration\cite{Aubert:2005exclisive}, is almost one
order-of-magnitude larger than the leading-order(LO)
predictions\cite{Braaten:2002fi,Liu:2002wq,Hagiwara:2003cw}. By
introducing the QCD perturbative
correction\cite{Zhang:2005ch,Gong:2007db}, and in combination with
relativistic correction\cite{He:2007te,Bodwin:2007ga}, this
discrepancy was largely resolved. Besides the challenges in the
exclusive process, the large ratio of
 $J/\psi$ production associated with charmed hadrons is also
 confusing, which was measured by Belle\cite{Abe:2002rb}
\begin{eqnarray}\label{eq:RccEX}
R_{c \bar c}=\frac{\sigma[e^+e^- \to J/\psi + c \bar c]}
{\sigma[e^+e^- \to J/\psi +X]} &=& 0.59 ^{+0.15}_{-0.13}\pm 0.12.
\end{eqnarray}
In contrast to the leading-order NRQCD predictions, this ratio is
only about 0.1\cite{Yuan:1996ep,cs,Braaten:1995ez}. The
next-leading-order(NLO) QCD corrections were also introduced in
$J/\psi$ inclusive production to resolve the discrepancy between
experimental measurements and LO calculations. The NLO QCD
corrections to $e^+e^-\rightarrow J/\psi c\bar{c}$ process enhance
the cross section with a $K$ factor of about
1.8\cite{Zhang:2006ay,Gong:2009ng}, and only about 20 percent for
$e^+e^-\rightarrow J/\psi gg$ process \cite{Ma:2008gq,Gong:2009kp}.
So the discrepancy is greatly alleviated. To check what role does
the color-octet process play, the NLO QCD corrections to color-octet
$J/\psi$ inclusive production was calculated\cite{Zhang:2009ym}.
Combining the relativistic corrections to $e^+e^-\rightarrow J/\psi
gg$\cite{He:2009uf}, it may imply that the values of color-octet
matrix elements are much smaller than the expected ones which are
estimated by using the naive velocity scaling rules.

Another interesting topic is the $e^+e^-\rightarrow
J/\psi+X_{non-c\bar c}$ production. Most recently, the prompt
$J/\psi$ production in association with charmed and non-charmed
final particles was measured\cite{Pakhlov:2009nj}
\begin{eqnarray}\label{eq:BelleSig}
\sigma(e^+e^-\rightarrow J/\psi+
X)&\hspace{-0.2cm}=&\hspace{-0.2cm}(1.17\pm 0.02\pm 0.07)pb,
 \nonumber \\
\sigma(e^+e^-\rightarrow J/\psi+c\bar
c)&\hspace{-0.2cm}=&\hspace{-0.2cm}(0.74\pm 0.08^{+0.09}_{-0.08})pb \nonumber \\
\sigma(e^+e^-\rightarrow J/\psi+X_{non-c\bar
c})&\hspace{-0.2cm}=&\hspace{-0.2cm}(0.43\pm 0.09\pm 0.09)pb.
\end{eqnarray}

These processes are investigated in Ref.\cite{He:2009uf,Jia:2009np},
the results show that including both the $O(\alpha_s)$ radiative
correction and the $O(v^2)$ relativistic correction, the
color-singlet contribution to $e^+e^-\rightarrow J/\psi gg$ has
saturated the latest observed cross section $e^+e^-\rightarrow
J/\psi+X_{non-c\bar c}$ measured by Belle.

Aside from the above QCD process, the pure QED process should also
be considered\cite{Chang:1998pz}. Especially, in our paper, we
calculate virtual-photon-associated production
$\sigma(e^+e^-\rightarrow J/\psi \gamma^*)\times
B(\gamma^*\rightarrow l\bar{l}(or \ Light \ Hadrons))$. The rest of
the paper is organized as follows. In Section
\uppercase\expandafter{\romannumeral 2}, we will give the
formulations of
 $e^+e^- \to J/\psi +\mu^+\mu^-$. In Section
\uppercase\expandafter{\romannumeral 3} , the QED production of
$e^+e^- \to J/\psi +LH$ is discussed. In section
\uppercase\expandafter{\romannumeral 4}, we will give the numerical
results and discussion. Finally we summarize our results in section
\uppercase\expandafter{\romannumeral 5}.


\section{The formulations  of $e^+e^- \to J/\psi +\mu^+\mu^-$}
In NRQCD factorization scheme, the cross section of  $e^+e^- \to
J/\psi +\mu^+\mu^-$ can be described as follows
\begin{eqnarray}
\label{amp2}
&& \hspace{-2.4cm}{\cal A}(e^+(k_1)e^- (k_2)\to  J/\psi (2p) +\mu^+(p_1)+\mu^-(p_2))\nonumber\\
&=&\sqrt{C_{S}} \sum_{L_{z} S_{z} }\sum_{s_1s_2}\sum_{jk}
\langle\frac 1 2 s_1;\frac 1 2 s_2\mid S S_{ z}\rangle
\langle L L_{ z };S S_{ z}\mid J J_{
z}\rangle\langle 3j;\bar{3}k\mid
1\rangle \times\nonumber\\
&& 
{\cal A}\big(e^+e^- \to
 c^{s1}_j(p)+\bar{c}^{s2}_k(p)+\mu^+(p_1)+\mu^-(p_2)\big)
\end{eqnarray}
where $\langle 3j ;\bar{3}k\mid1\rangle =1/\sqrt 3$,
$\langle s_1;s_2\mid S S_{ z}\rangle$, $\langle L
L_{ z };S S_{ z}\mid J J_{ z}\rangle$ are respectively the color-SU(3),
spin-SU(2), and angular momentum Clebsch-Gordan coefficients for
$c\bar{c}$ pairs projecting out appropriate bound states. ${\cal A}(e^+e^- \to
 c_j(p)+\bar{c}_k(p)+\mu^+(p_1)+\mu^-(p_2))$ is the scattering
amplitude for $c \bar c$ production.  The coefficient $C_{S}$ can be
related to the radial wave function of the bound state and reads
\begin{equation}
\label{eq:csRs0}
C_{S}=\frac{1}{4\pi}\mid R_{S}(0) \mid^{2}.
\end{equation}

We introduce the spin projection operators $P_{SS_z}(p,q)$ as
\begin{equation}
P_{SS_z}(p,q)\equiv\sum\limits_{s_1s_2 }\langle s_1;s_2|SS_z\rangle
v(p-q;s_1)\bar{u}(p+q;s_2).
\end{equation}
Expanding $P_{SS_z}(p,q)$ in terms of the relative momentum $q$, we
get the projection operators at leading term of $q$, which will be
used in our calculation, as follows
\begin{eqnarray}
\label{pjs} P_{1S_z}(p,0)&=&\frac{1}{\sqrt{2}}\ \epsilon\!\!
/^*(S_z)(p\!\!/+M/2).
\end{eqnarray}
where $M$ is the mass of the charmonium. It is two times of charm
quark mass $m$ in the non-relativistic approximation. The polarized
cross section can be calculated by defining the longitudinal
polarization vector as follows
\begin{equation}
\epsilon^{\mu}_L(p)=\frac{2p^{\mu}}{M}-\frac{Mn^{\mu}}{2n\cdot p},
\end{equation}
where $4p^2=M^2$ and $n^{\mu}=(1,-\vec{p}/\mid \vec{p} \mid)$.

\begin{figure}
\begin{center}
\includegraphics[width=8.5cm]{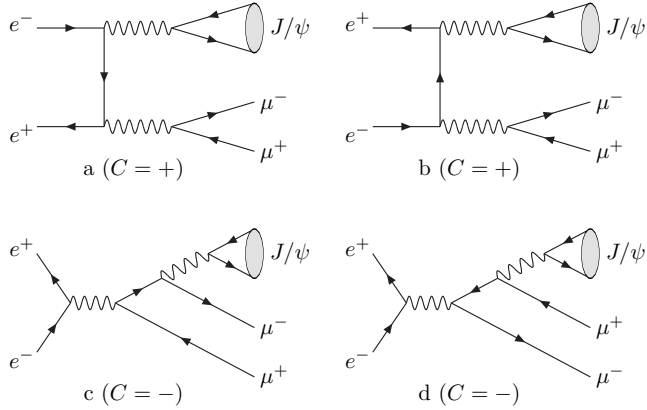}
\caption{\label{fig:eeJpsiLLFeyn} The Feynman diagrams of $e^+e^-\to
J/\psi \mu^+\mu^-$. }
\end{center}
\end{figure}

The Feynman diagrams  of $e^+e^-\to J/\psi \mu^+\mu^-$ are shown in
Fig.\ref{fig:eeJpsiLLFeyn}. The process of final states with plus
charge parity is depicted in diagram Fig.\ref{fig:eeJpsiLLFeyn}$(a,
b)$ and denoted as $C=+$. The process of final states with minus
charge parity is depicted in diagram Fig.\ref{fig:eeJpsiLLFeyn}$(c,
d)$ and denoted as $C=-$. The $C=+$ process is dominant, and the
$C=-$ process is suppressed by a factor of
\begin{eqnarray}\label{eq:FSc}
f \sim \ln \left(\frac{s}{4 m_\mu^2}\right)\frac{s}{M^2},
\end{eqnarray}
here the logarithm term come from quasi-collinear divergence with
$m_\mu^2 \ll s$. The $s/M^2$ term come the photon propagator. One
can get $f \sim 10^{-2}$ for $e^+e^-\to J/\psi \mu^+\mu^-$ at $\sqrt
s =10.6$~GeV for $B$ factories.

\section{The
QED production of $e^+e^- \to J/\psi + LH$}

Similar to the $e^+e^-\to J/\psi \mu^+\mu^-$ process, the fragment
process is also dominant in $e^+e^- \to J/\psi +Light \ Hadrons$
process. The fragment process can be considered as $e^+e^-\to
\gamma^* \gamma^*$, then the virtual photons fragment into $J/\psi$
and light hadrons respectively. By using the approach of the
calculation of the hadronic part of the muon
$g-2$\cite{Ezhela:2003pp}, this process can be described as
\begin{eqnarray}\label{eq:LHest}
\frac {{\rm d}\sigma^{QED}[e^+e^- \to J/\psi + LH]}{{\rm d }m^2_{LH}} \sim \left.\frac
{{\rm d}\sigma[e^+e^- \to J/\psi + \mu^+\mu^-]}{{\rm d }m^2_{\mu^+\mu^-}}
\times R^{had}(m^2_{\mu^+\mu^-})\right|_{m_{\mu^+\mu^-}=m_{LH}},
\end{eqnarray}
where
\begin{eqnarray}\label{eq:Rhad}
R^{had}(\Lambda^2)=\left.\frac {\sigma[e^+e^- \to
hadrons]}{\sigma[e^+e^- \to
\mu^+\mu^-]}\right|_{m^2_{e^+e^-}=\Lambda^2},
\end{eqnarray}
because of the contribution of the $C=-$ process is negligible,
after subtracting the effect of $\gamma^* \to c \bar c$ from
$R^{had}(\Lambda^2)$ by a naive factor $4/3 \Theta(\Lambda -
M_{c\bar c-The})$, we get
\begin{eqnarray}\label{eq:LHestNOccbar}
&&\frac {{\rm d}\sigma^{QED}[e^+e^- \to J/\psi + LH]}{{\rm d }m^2_{LH}} \nonumber \\
 &=& \left.\frac {{\rm d}\sigma[e^+e^- \to J/\psi + \mu^+\mu^-]}{{\rm d }
 m^2_{\mu^+\mu^-}}\times \left[R^{had}(m^2_{\mu^+\mu^-})-\frac 4 3 \Theta(m^2_{\mu^+\mu^-} - M^2_{c\bar c-The})\right]\right|_{m_{\mu^+\mu^-}=m_{LH}},
\end{eqnarray}
where $\Theta$ is step function, $M_{c\bar c-The}$ should be
correspond to the $c \bar c$ threshold of $M_{J/\psi}$, $2M_D$, etc.
The uncertainties should be discussed in the next section.

\section{Numerical result}

In numerical calculations, the parameters are selected as
\cite{Amsler:2008zzb}:
\begin{eqnarray}
M_\mu&=&0.1057GeV,\ \ \  M_{J/\psi}=3.0969GeV,  \ \ \ \sqrt s=10.6GeV, \nonumber \\
M_\tau&=&1.7768GeV,\ \ \ M_{\psi(2S)}=3.686GeV,\ \ \  \alpha=1/132.33
\end{eqnarray}

The table of $R^{had}(\lambda^2)$ have been given in
Ref.\cite{Ezhela:2003pp}. We construct an interpolation of
$R^{had}(\Lambda^2)$ corresponding to the table with first-order
interpolation and setting $R^{had}(\Lambda^2)=0$ when $\Lambda$ is
less than $\lambda_{min}$ in the table.

The wave function at the origin can be extracted from the leptonic
width $\Gamma (V\rightarrow l^{+}l^{-})$
\begin{equation}
|R_S(0)|^2=\frac{m^2_{V}}{4 e_Q^2\alpha^2}\Gamma[V \to e^+e^-] .\label{eq:rs0VLL}
\end{equation}

The leptonic width of charmonium decays into $e^+e^-$ has been given
in Ref.\cite{Amsler:2008zzb}
\begin{eqnarray}\label{eq:PDGVll}
\Gamma[J/\psi \to e^+e^-]&=&5.55 \pm 0.14 keV, \nonumber \\
\Gamma[\psi(2S) \to e^+e^-]&=&2.38 \pm 0.04 keV.
\end{eqnarray}

When we calculate the prompt production cross sections of  $J/\psi$,
we take into account  the feeddown contribution from $\psi(2S)$ by
$B[\psi(2S) \to J/\psi+X]=(57.4 \pm 0.9)\%$\cite{Amsler:2008zzb} and
ignore the contribution of the other charmonium. Then we can get the
direct production cross section of  $ J/\psi(\psi (2S))$ associated
with $\tau^+\tau^-$ and $\mu^+\mu^-$ at $B$ factories as
\begin{eqnarray}
\sigma^{direct}[e^+e^- \to J/\psi +\mu^+\mu^- ]&=&56 \pm 2\ fb \nonumber \\
\sigma^{direct}[e^+e^- \to J/\psi+ \tau^+\tau^- ]&=&6.4 \pm 0.2\  fb
\end{eqnarray}
and
\begin{eqnarray}
\sigma^{direct}[e^+e^- \to \psi (2S) +\mu^+\mu^- ]&=&20 \pm 1\ fb \nonumber \\
\sigma^{direct}[e^+e^- \to \psi (2S)+ \tau^+\tau^- ]&=&1.8 \pm 0.1\  fb
\end{eqnarray}

Most of the uncertainties come from the uncertainty of leptonic
width in Eq.(\ref{eq:PDGVll}). The others come from the effect of
fine structure constant $\alpha$ and higher order QED corrections
and so on. The QCD corrections have been taken into account in the
leptonic width of $J/\psi (\psi(2S))$.

The cross sections for $C=-$ process is only $1.6\%(1.0\%)$ of that
for $C=+$ process in $J/\psi(\psi (2S))$ production associated with
$\mu^+\mu^-$. And the ratio is about  $6.0\%(3.9\%)$ in $J/\psi(\psi
(2S))$ production associated with $\tau^+\tau^-$. These results are
in agreement with the estimation in Eq.(\ref{eq:FSc}). So the
contribution of $C=-$ process can be ignored in the calculation of
electromagnetic $J/\psi(\psi(2S))$ production associated with light
hadrons. Finally, we get the direct production cross section of $
J/\psi(\psi (2S))$ associated with light hadrons at $B$ factories as
\begin{eqnarray}
\sigma_{QED}^{direct}[e^+e^- \to J/\psi +LH ]&=&100 \pm 5\ fb \nonumber \\
\sigma_{QED}^{direct}[e^+e^- \to \psi(2S)+ LH]&=&36 \pm 1 \  fb
\end{eqnarray}
here we choose $M_{c\bar c-The}=2M_D$ in Eq.(\ref{eq:LHestNOccbar}).
If we choose $M_{c\bar c-The}=M_{J/\psi}$, there is a difference of
$-1fb$. So the uncertainties from $M_{c\bar c-The}$ can be ignored.
Most of the uncertainties come from $R^{had}$ and the leptonic decay
width.

\begin{figure}
\begin{center}
\includegraphics[width=8.5cm]{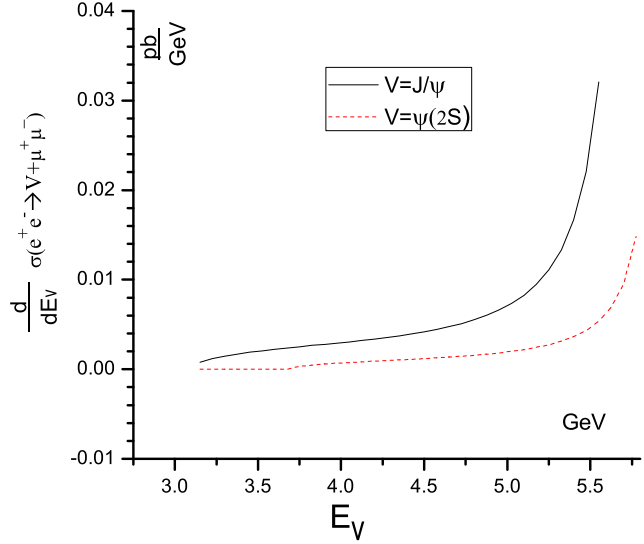}
\caption{\label{fig:muP50} The $J/\psi$ and $\psi(2S)$ energy
spectra of direct production processes $e^+e^-\to V+ \mu^+\mu^-\ \
(V=J/\psi,\psi(2S))$. }
\end{center}
\end{figure}

\begin{figure}
\begin{center}
\includegraphics[width=8.5cm]{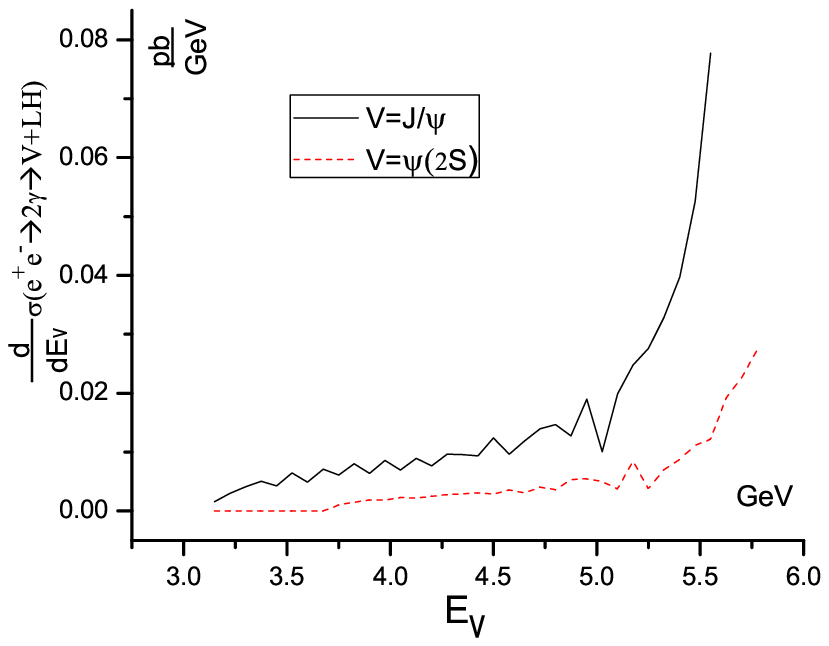}
\caption{\label{fig:LHP50} The $J/\psi$ and $\psi(2S)$ energy
spectra  of direct production processes  $e^+e^-\to 2 \gamma \to V +
LH\ \ (V=J/\psi,\psi(2S))$. }
\end{center}
\end{figure}

\begin{figure}
\begin{center}
\includegraphics[width=8.5cm]{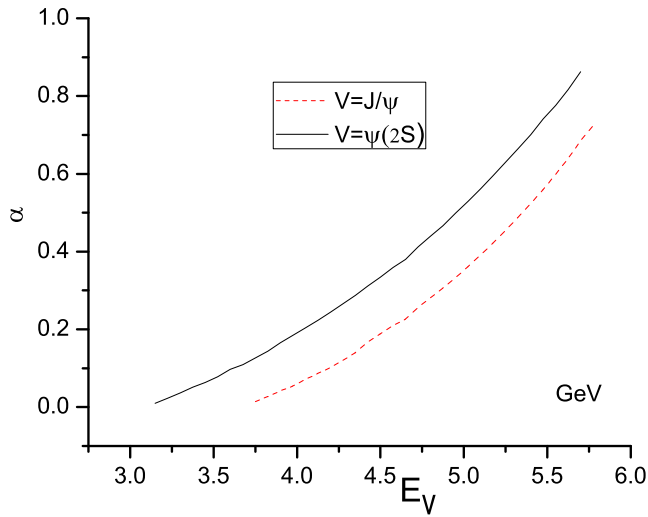}
\caption{\label{fig:alpha} The helicities of $J/\psi$ and $\psi(2S)$
direct production processes $e^+e^-\to V+ \mu^+\mu^-\ \
(V=J/\psi,\psi(2S))$ as a function of the energy of $J/\psi$ and
$\psi(2S)$.}
\end{center}
\end{figure}

\begin{figure}
\begin{center}
\includegraphics[width=8.5cm]{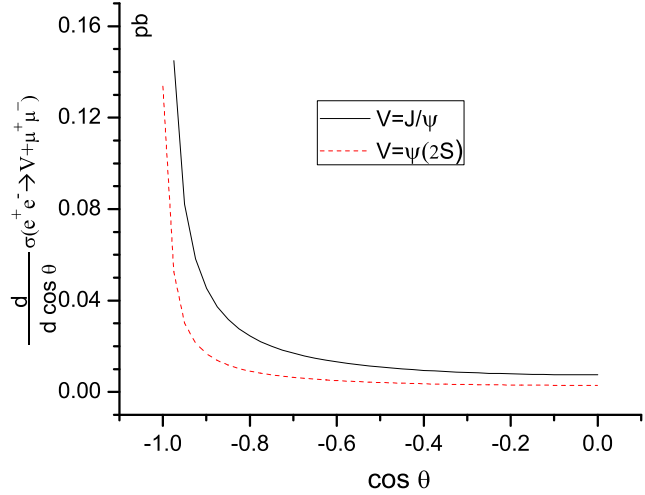}
\caption{\label{fig:muTht} The angular distributions  of direct
production processes $e^+e^-\to  V+ \mu^+\mu^-(V=J/\psi,\psi(2S))$.
Here $\theta$ is the angle between $J/\psi(\psi(2S))$ momentum and
beam.}
\end{center}
\end{figure}

\begin{figure}
\begin{center}
\includegraphics[width=8.5cm]{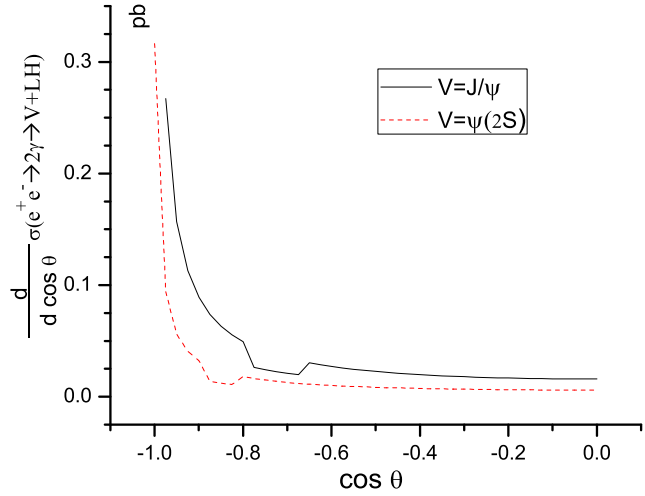}
\caption{\label{fig:LHTht} The angular distributions  of direct
production processes $e^+e^-\to 2 \gamma \to V + LH\ \
(V=J/\psi,\psi(2S))$. Here $\theta$ is the angle between
$J/\psi(\psi(2S))$ and beam.}
\end{center}
\end{figure}

The energy distributions of direct $J/\psi(\psi (2S))$ production
from the processes $e^+ e^- \rightarrow J/\psi(\psi (2S) l\bar{l}$
and $e^+ e^- \rightarrow J/\psi(\psi (2S)) L H$ are shown in
Fig.\ref{fig:muP50} and Fig.\ref{fig:LHP50}, respectively.
Unfortunately, the endpoint peak was not measured in
Ref.\cite{Pakhlov:2009nj}. The polarization of $J/\psi(\psi(2S))$
direct production for $e^+e^-\to V+ \mu^+\mu^-\ \
(V=J/\psi,\psi(2S))$ process as a function of the energy of $J/\psi$
is shown in Fig.\ref{fig:alpha}. The polarization of $J/\psi$
production associated with light hadrons is similar to the results
of $e^+e^-\to V+ \mu^+\mu^-$ process. The angular distributions of
direct $J/\psi(\psi (2S))$ production from the processes $e^+e^-\to
V+ \mu^+\mu^-(V=J/\psi,\psi(2S))$ and $e^+e^-\to 2 \gamma* \to V +
LH\ \ (V=J/\psi,\psi(2S))$ are shown in Fig.\ref{fig:muTht} and
Fig.\ref{fig:LHTht}, respectively.

%

 Now we give the prompt production cross sections of electromagnetic
 $J/\psi$ production associated with leptonic pairs and light hadrons as
\begin{eqnarray}
\sigma_{QED}^{prompt}[e^+e^- \to J/\psi +\mu^+\mu^- +X]&=&68 \pm 2 \  fb\nonumber \\
\sigma_{QED}^{prompt}[e^+e^- \to J/\psi +\tau^+\tau^- +X]&=&7.4 \pm 0.1 \  fb\nonumber \\
\sigma_{QED}^{prompt}[e^+e^- \to J/\psi+ LH]&=&121 \pm 5\  fb.
\end{eqnarray}


\section{Summary}

In summary,  the electromagnetic productions of $J / \psi$
associated with light hadrons(LH) and leptonic pairs($\mu^+\mu^-$,
$\tau^+\tau^-$) at $B$ factories are studied. We find that the
direct electromagnetic production cross sections of $J /
\psi$($\psi(2S)$) associated with light hadrons is about
$0.10(0.04)$~pb. The direct production cross section of $J /
\psi$($\psi(2S)$) associated with $\mu^+\mu^-$ is about
$0.056(0.020)$~pb. If we include the contribution from $\psi(2S)$
decay, we can get the prompt cross section $\sigma[e^+e^- \to J/\psi
+\mu^+\mu^-+X]=(68 \pm 2)$~pb, about $(16 \pm 5)\%$ of the Belle
data $\sigma[e^+e^- \to J/\psi +X_{non-c\bar c}]=(0.43\pm 0.09\pm
0.09)$~pb, meanwhile the $e^+e^- \to J/\psi +\tau^+\tau^-$
 process only contributes $2 \%$. The prompt cross section $\sigma[e^+e^-
\to J/\psi +Light\ Hadrons]=(0.121 \pm 5)$~fb is  about $(28 \pm
8)\%$ of the Belle data. Unfortunately, the endpoint peak of energy
distribution for $J/\psi$ electromagnetic production associated with
leptonic pairs and light hadrons was not measured in
Ref.\cite{Pakhlov:2009nj}. The polarization of $J/\psi$
electromagnetic production associated with light hadrons is
transversal, while the polarization of $J/\psi$ inclusive production
associated with light hadrons from QCD process is
longitudinal\cite{Jia:2009np}. We also notify that the charge parity
of final states is plus for the QED process calculated in our paper.
And it is minus for color singlet process $e^+e^- \to J/\psi +gg$
and color octet process $e^+e^- \to J/\psi +g$.

\begin{acknowledgments}
The authors would like to thank Professor K.T.Chao for useful
discussion. This work was supported by the National Natural Science
Foundation of China (No 10805002, No 10875055), FANEDD2007B18, the
Project of Knowledge Innovation Program (PKIP) of Chinese Academy of
Sciences, Grant No. KJCX2.YW.W10 and the Education Ministry of
Liaoning Province.

\end{acknowledgments}


\providecommand{\href}[2]{#2}\begingroup\raggedright\endgroup
\end{document}